# Prediction of Orientational Phase Transition in Boron Carbide

M. Widom[a] (widom@andrew.cmu.edu) and W. P. Huhn[a] (wph@andrew.cmu.edu)
[a]Department of Physics, Carnegie Mellon University
5000 Forbes Avenue
Pittsburgh, PA  15213
United States of America
412-268-7645

Corresponding Author:  M. Widom[a]

**Abstract:** The assessed binary phase diagram of boron-carbon exhibits a single alloy phase designated "$B_4C$" with rhombohedral symmetry occupying a broad composition range that falls just short of the nominal carbon content of 20%. As this composition range is nearly temperature independent, the phase diagram suggests a violation of the third law of thermodynamics, which typically requires compounds to achieve a definite stoichiometry at low temperatures. By means of first principles total energy calculations we predict the existence of two stoichiometric phases at $T$=0K: one of composition $B_4C$ with monoclinic symmetry; the other of composition $B_{13}C_2$ with rhombohedral symmetry. Using statistical mechanics to extend to finite temperatures, we demonstrate that the monoclinic phase reverts to the rhombohedral phase above $T$=600K, along with a slight reduction on carbon content.

## 1. Introduction

Only a single compound is reported in the boron-carbon binary system, the nonstoichiometric compound boron carbide, often named "$B_4C$" whose composition range extends from 9% to 19.2% carbon. Notice this range excludes the 20% carbon that is expected for the implied stoichiometry of $B_4C$. Presumably the composition range is due to substitutional disorder between B and C atoms. As commonly drawn, the composition range is nearly independent of temperature over the range 1000-2000 C (1273-2273 K) [1, 2]. If this composition range were to persist at low temperatures it would imply the existence of substitutional disorder, and hence configurational entropy, at $T$=0K, in apparent violation of the third law of thermodynamics [3, 4]. Other phase diagrams have been proposed but rejected, including one [5, 6] containing a large number of distinct phases, including a phase labeled $B_{13}C_2$ with variable stoichiometry at high temperature but reaching a fixed stoichiometry as T approaches 0K.

Boron carbide displays rhombohedral symmetry (space group $R\bar{3}m$, #166), based on a 15 atom primitive cell (Pearson symbol hR15). Its structure is similar to α-boron (Pearson hR12), which places 12-atom icosahedra at vertices of the rhombohedron, but boron carbide includes an additional chain of three atoms in the center of the rhombohedral cell aligned along the 3-fold axis (see Fig. 1). The symmetry of the structure depends on the placement of chemical species B and C among these sites. Maintaining full boron occupancy ($B_{12}$) of the icosahedron and placing a C-B-C chain along the axis results in a rhombohedral structure [7] of composition $B_{13}C_2$ (carbon fraction $x_C$=0.133). Such structures have been confirmed experimentally [8]. At composition $B_{12}C_3=B_4C$, rhombohedral symmetry is maintained with a $B_{12}$ icosahedron and a C-C-C chain [7, 9, 10]. However, first principles calculations [11, 12] suggest total energy is reduced by substituting carbon for one icosahedral boron atom at the "polar" position indicated in Fig. 1 (right). The symmetry breaking caused by the chemical substitution of carbon onto the icosahedron reduces the symmetry from rhombohedral to monoclinic, yielding Pearson type mC30 (space group Cm, #8).

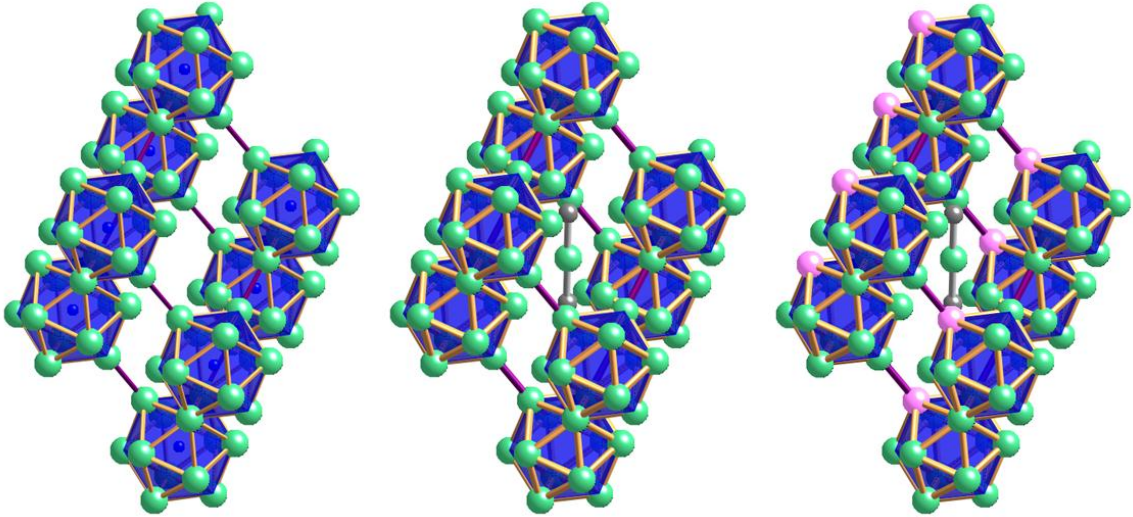

Figure 1. (left) α-boron (hR12); (center) $B_{13}C_2$ (hR15); (right) $B_4C$ (mC30). Boron atoms in green, carbon in gray. Pink atom labels "polar" carbon site.

The higher rhombohedral symmetry can be restored, at any carbon content, by randomly substituting carbon among all six polar icosahedral positions with equal weight. Indeed, experimental evidence [13] suggests the true structure is a composition-dependent mixture of $B_{12}$ and $B_{11}C_1$ icosahedra, and likewise a mixture of C-B-C, C-B-B and B-V-B chains (V=vacancy).

We carry out first principles total energy calculations to study the distribution of carbon and boron atoms on the underlying hR15 crystal structure. The immediate output of our calculations is a predicted phase diagram in the limit of $T$=0K that reveals two stable stoichiometric compounds – precisely the $B_{13}C_2$ and $B_4C$ structures illustrated in Fig. 1. Identical conclusions were previously reached by Bylander and Kleinman [14] and by Vast [15]. However, we explore the ensemble of higher energy structures obtained by varying the placement of carbon atoms among the sites, both of the 15 atom primitive cell and of various supercells. We then calculate the configurational partition function in order to obtain the finite temperature free energy. Our calculational methods are discussed in the following section.

This free energy predicts that the $B_4C$ phase is unstable above $T$=600K. The principle excitation that destabilizes the $B_4C$ phase corresponds to independent rotations of the $B_{11}C_1$ icosahedra in adjoining primitive cells, resulting in a restoration of full rhombohedral symmetry. A secondary effect is the replacement of polar carbons with boron, causing a slight reduction in the carbon content below the ideal $B_4C$ stoichiometry. Because the $B_{13}C_2$ phase already has rhombohedral symmetry, it suffers no phase change as temperature rises, until melting around $T$=2800K. However, the composition of "$B_{13}C_2$" becomes highly variable as temperature rises. Indeed, the high temperature limit of the phase termed "$B_4C$" should properly be considered as the "$B_{13}C_2$" phase at a high C content. Note the confusion that arises from naming phases according to their approximate stoichiometry.

## 2. Theory/calculations

### 2.1 First principles total energy calculations

First principles total energy calculations yield the enthalpy of formation of specific trial structures. Enthalpy minima, i.e. the convex hull of $H(x)$ where $x$ is composition, correspond to predicted low temperature stable phases. Structures lying above the convex hull are unstable to decomposition into competing stable phases at low temperature, but they can contribute significantly to the free energy at elevated temperatures as outlined in the following section.

Our total energy calculations are based on electronic density functional theory, the highest level of theory presently available for extended structures. We utilize projector augmented wave (PAW) potentials [16], an all electron generalization of pseudopotentials, as implemented in the plane wave code VASP [17, 18]. For our exchange correlation potential we choose the PW91 generalized gradient approximation (GGA) [19]. All structures are fully relaxed in both atomic and cell coordinates, yielding enthalpy $H_i$ for structure $i$ at $T=P=0$. Default energy cutoffs of 400 eV were used for plane wave calculations. Brillouin zone integrations were performed using Monkhorst-Pack $k$-point meshes [20] sufficiently dense that energies had converged to within 0.001 eV/atom. We subtract all enthalpies from the tie-line connecting elemental boron (β-boron, Pearson hR141 optimized with 107 atoms [21]) with elemental carbon (graphite, Pearson hP4), yielding the enthalpy of formation $\Delta H_i$. We also define the energy of instability, $\Delta E_i$, as the energy of an unstable structure above the tie-lines joining competing stable phases on the convex hull of $H(x)$.

In order to fully explore configuration space we carry out calculations both in the primitive cell and in super cells, including a 2x1x1 cell doubled along the rhombohedral $a$ axis, the hexagonal unit cell formed from three primitive cells, and a 2x2x1 super cell doubled along both the $a$ and $b$ rhombohedral axes containing four primitive cells. We concentrate our efforts on structures expected to be low in energy based on results from the primitive cell calculations. For the primitive cell, we used all possible symmetry independent primitive cell configurations up to 27% carbon concentration. In supercells we explored many (but not all) of the possible combinations of primitive cell configurations. The vast majority of combinations yield high energy. Configurations containing the well-known structure elements of $B_{12}$, $B_{11}C_1$, and $B_{10}C_2$ icosahedra, and containing C-B-C and C-B-B chains prove reasonably low in energy. The combination of a $B_{10}C_2$ icosahedron with a $B_{12}$ icosahedron forms the "bipolar defect" [22]. All possible positions of carbon atoms within each icosahedron and chain are considered. Structures are grouped into symmetry-equivalent classes and we record their multiplicities, that is, the number of distinct symmetry-equivalent structures. Enthalpies of selected structures are shown in Fig. 2 for each cell size considered, and listed in Table 1 for the hexagonal unit cell.

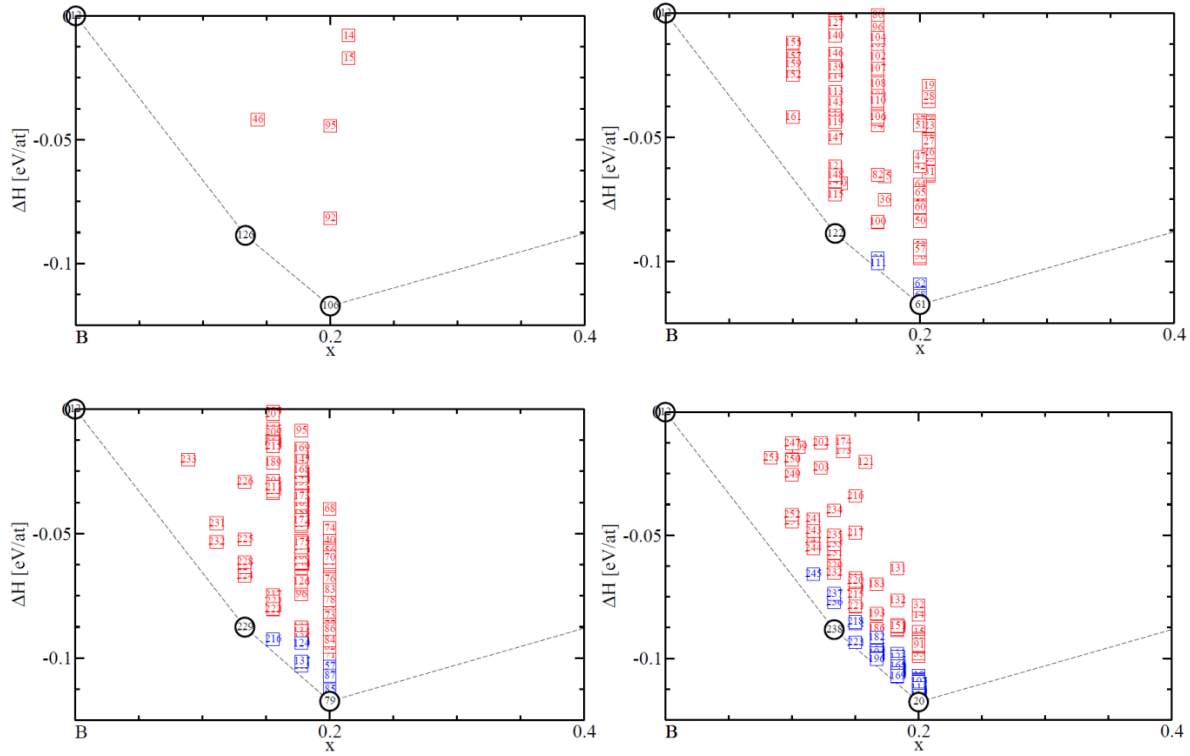

Figure 2. Cohesive energies of boron carbide structures, left to right, top to bottom: (a) 15 atom primitive cell; (b) 30 atom 2x1x1 super cell; (c) 45 atom hexagonal unit cell; (d) 60 atoms 2x2x1 super cell. Heavy black circles at x=0.133 and 0.20, respectively, correspond to the ideal $B_{13}C_2$ and $B_4C$ structures displayed in Fig. 1. Data points in blue have $0<\Delta E<0.015$ eV/atom, while red has $0.015<\Delta E$.

| Name | $\Delta E$ (meV/atom) | $\Delta H$ (meV/atom) | $x_C$ | Cell content | $\Omega$ | Description |
|---|---|---|---|---|---|---|
| C.hP4 | 0 | 0 | 1 | $C_4$ | 1 | Graphite |
| B.hR141 | 0 | 0 | 0 | $B_{107}$ | 1 | β Boron |
| (p0tt')(p0tt')(p0tt') | 0 | -117.4 | 0.2 | $B_{36}C_9$ | 6 | $B_4C$ (monoclinic) |
| (tt')(tt')(tt') | 0 | -87.6 | 0.1333 | $B_{39}C_6$ | 1 | $B_{13}C_2$ (rhombohedral) |
| (p0tt')(p0tt')(p1tt') | 3.0 | -114.4 | 0.2 | $B_{36}C_9$ | 36 | Rotation |
| (p0tt')(p0tt')(tt') | 4.4 | -103.1 | 0.1778 | $B_{37}C_8$ | 18 | Mixed icosahedra |
| (p0tt')(p1tt')(p2tt') | 4.5 | -112.9 | 0.2 | $B_{36}C_9$ | 12 | Rotation |
| (p0tt')(tt')(tt') | 5.2 | -92.4 | 0.1556 | $B_{38}C_7$ | 18 | Mixed icosahedra |
| (p0tt')(p1tt')(tt') | 6.2 | -101.3 | 0.1778 | $B_{37}C_8$ | 36 | Mixed icosahedra |
| (p1p1'tt')(tt')(p0tt') | 10.6 | -106.8 | 0.2 | $B_{36}B_9$ | 36 | Bipolar defect |
| (p0tt')(p1'tt')(tt') | 13.4 | -94.0 | 0.1778 | $B_{37}C_8$ | 18 | Mixed icosahedra |
| (e0tt')(p0'tt')(p0tt') | 13.9 | -103.5 | 0.2 | $B_{36}B_9$ | 18 | One equatorial, rotation |

Table 1. Table of low $\Delta E$ structures for pure elements and the hexagonal boron carbide cell. In our naming scheme, adapted from the primitive cell, sites listed are those containing carbons. t and t' denotes the two ends of the chain (so that tt' denotes a C-B-C chain), p{0,1,2} denotes the polar atoms on the "north" side of the icosahedra, p'{0,1,2} denotes the polar atoms on the opposite ("south") side of the icosahedra. Likewise e{0,1,2} and e'{0,1,2} denotes the atoms just to the north and south of the equator. $\Omega$ is the multiplicity.

## 2.2 Configurational free energy

To go from total energy of a configuration of $N=N_B+N_C$ atoms $E(\{r_j\})$ in volume $V$ to free energy, we must calculate the statistical mechanical partition function

$$Z(N_B, N_C, V, T) = \int_V \prod_j dr_j\, e^{-E(\{r_j\})/k_BT}.$$

However, we prefer to work at fixed pressure, hence we switch to the Gibbs $(N,P,T)$ ensemble,

$$Q(N_B, N_C, P, T) = \int dV\, e^{-PV/k_BT}\, Z(N_B, N_C, V, T).$$

Since the dominant contributions to the partition function arise at relatively low energies, we can replace the integral over positions $\{r_j\}$ and volumes $V$ with a sum over local energy minima $i$ of a Boltzmann-like factor $exp\left(-F_{vib}^{(i)}/k_BT\right)$ related to the vibrational free energy of the local minimum $i$,

$$Q(N_B, N_C, P, T) \approx \sum_i e^{-F_{vib}^{(i)}/k_BT}.$$

As a further approximation, we replace $F_{vib}^{(i)}$ with the enthalpy $\Delta H_i$. While this assumption is certainly not valid, the result should affect our conclusions only quantitatively, not qualitatively, provided the vibrational properties of different relaxed configurations differ only slightly. Taking the logarithm, we obtain the Gibbs free energy

$$G(N_B, N_C, P, T) = -k_BT \ln Q(N_B, N_C, P, T) = \mu_B N_B + \mu_C N_C.$$

Since the total number of atoms $N=N_B+N_C$ is fixed, for a given primitive cell or super cell, it is convenient to re-express $G(N_B, N_C, T) = \mu_B N + \delta\mu N_C$, where $\delta\mu = \mu_C - \mu_B$. In fact, we can change to a semi-grand canonical potential

$$\Sigma(N, \delta\mu, T) = \sum_{N_C} e^{\delta\mu N_C/k_BT}\, Q(N-N_C, N_C, T)$$

and use the chemical potential difference to control the composition $x_C = N_C/(N_B+N_C)$. The free energy in this ensemble is simply $Y(N, \delta\mu, T) = -k_BT \ln Z = \mu_B N$. Inspecting the enthalpies shown in Fig. 2 we see that choosing $\delta\mu=0$ at low temperature places us in the $B_4C$ phase at $x_C=0.200$, while $\delta\mu=-0.55$ corresponds to $B_{13}C_2$ at $x_C=0.133$.

## 3. Results and Discussion

Differentiating $Y$ with respect to $\delta\mu$ gives the number of carbon atoms $N_C$, while two derivatives with respect to $T$ give the heat capacity

$$C_p(N, \delta\mu, T) = \frac{d^2Y}{dT^2} = -k_B \frac{d^2(T\ln Z)}{dT^2} = \frac{\langle H^2\rangle - \langle H\rangle^2}{k_BT^2}.$$

Where angle brackets indicate the configurational ensemble average, and $H$ is the enthalpy of formation. Results for $C_P$ are shown in Fig. 3, using the energies displayed in Fig. 2. Note that we only use subsets of the possible energies (those in black and blue), chosen to accurately reflect the thermodynamics up to $T=1000$K but omitting higher energy states that would cause the heat capacity to rise again as $T$ increases. These states were omitted for reasons of limitations in computer time when dealing with larger cells and have no impact on the curves shown in Fig. 3.

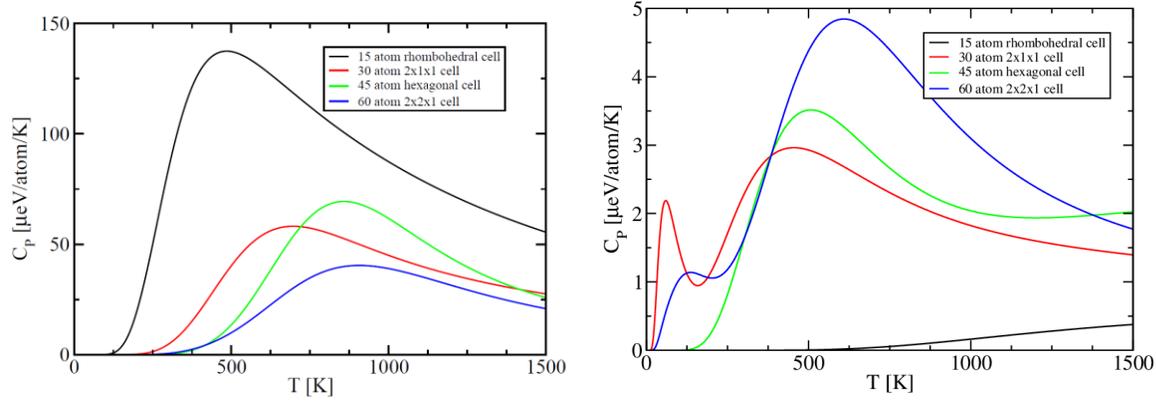

Figure 3. Heat capacity at $\delta\mu = -0.55$ (left) and $\delta\mu = 0$ (right).

A phase transition is clearly indicated at $\delta\mu = 0$ by the peak in heat capacity around $T=500-600$K. The peak is growing in strength as system size increases, indicating a divergence in the infinite size limit. The peak is entirely absent in the case of the single primitive cell because the relevant excitation, rotations of C among polar B sites, is an exact symmetry of the primitive cell. In fact, there is a 6-fold degeneracy of the ground state. This degeneracy has no consequence in the thermodynamic limit as it is nonextensive. A weak peak is observed around $T=100-200$K in the two even super cells (30 and 60 atoms). These are due to a loss of inversion symmetry but with non-extensive multiplicity, so that the peak height tends to zero in the thermodynamic limit. For super cells of two or more primitive cells, the primary peak corresponds to rotation of the carbon atoms among icosahedral polar sites. The relevant excitations are rotational degrees of freedom in the selection of polar site on which to place carbon atoms. Because this phase transition corresponds to unlocking the orientations of polar carbons, rhombohedral symmetry is restored in the high temperature phase.

Although there is a heat capacity peak at $\delta\mu = -0.55$, this peak diminishes in magnitude as system size increases, and thus is not evidence of a phase transition. It is caused by excitation of $B_{11}C_1$ icosahedra, while only $B_{12}$ exist in the ground state at this chemical potential. Thus the $B_{13}C_2$ phase persists to high temperature with no change in thermodynamic state. The presence of polar carbons as low energy excitations allows this phase to exhibit a broad composition range. Owing to orientational disorder, full rhombohedral symmetry is present throughout the entire temperature and composition range.

By varying the chemical potential δµ we may vary the compositions of the $B_4C$ and $B_{13}C_2$ phases. We find that the composition of $B_4C$ remains nearly fixed at $x_C=1/5=0.200$ at low temperatures. $B_{13}C_2$ shows some variability around the value $x_C=2/15=0.133$ at low temperature, and high variability at high temperatures, reaching a low value of $x_C=0.070$ a high value of $x_C=0.198$. We do not expect our low carbon limit to be accurate because we have not extensively studied boron-rich structures. The carbon content decreases at high temperature owing to the entropy of selecting icosahedra on which to replace polar carbon atoms with boron. The high temperature limit at δµ=0 is precisely the same thermodynamic phase as was obtained at δµ=-0.55.

## 4. Conclusions

In conclusion, our first principles calculations predict the existence of two distinct boron carbide phases with differing symmetries at low temperatures. One has precise stoichiometry $B_{13}C_2$ and rhombohedral symmetry. The other has precise stoichiometry $B_4C$ and monoclinic symmetry. At high temperatures the $B_4C$ phase undergoes an orientational disordering transition and becomes thermodynamically indistinct from the rhombohedral $B_{13}C_2$ phase.

## 4. Acknowledgments

We thank Nathalie Vast, Helmut Werheit and Koun Shirai for useful discussions.